\documentclass[12pt,a4paper]{article}
\usepackage{a4wide}
\usepackage{amsmath,amsthm,amsfonts,amssymb,bbm}
\usepackage{graphicx,psfrag}
\usepackage{cite}

\numberwithin{equation}{section}

\newtheorem{prop}{Proposition}
\newtheorem{thm}[prop]{Theorem}
\newtheorem{lem}[prop]{Lemma}

\newenvironment{proofOF}[2]{\removelastskip\vspace{6pt}\noindent {\it Proof of #1.}~\rm#2}{\qed \par\vspace{6pt}}

\title{Current moments of 1D ASEP by duality}
\author{Takashi Imamura\thanks{University of Tokyo,
e-mail: imamura@jamology.rcast.u-tokyo.ac.jp},
Tomohiro Sasamoto\thanks{Chiba University, sasamoto@math.s.chiba-u.ac.jp}
}

\date{\today}

\begin{document}
\maketitle \sloppy

\begin{abstract}
We consider the exponential moments of integrated currents 
of 1D asymmetric simple exclusion process using the duality 
found by Sch\"utz. For the ASEP on the infinite lattice 
we show that the $n$th moment is reduced to the problem of 
the ASEP with less than or equal to $n$ particles.  
\end{abstract}

\section{Introduction}\label{intro}

The one dimensional asymmetric simple exclusion process (ASEP)
is a many-particle stochastic process in which each particle 
is an asymmetric random walker but with exclusion interaction 
among particles\cite{Liggett1985,Liggett1999}. We consider the 
ASEP with hopping rate $p$ to the right and $q$ to the left 
with $p+q=1$. The ratio is denoted by $\tau=p/q$.

The ASEP can be defined either on a finite lattice or on an 
infinite lattice. It is of much current interest to study 
fluctuation properties of the ASEP on $\mathbb{Z}$ because for 
this case one can perform detailed analysis using the 
connection to random matrix theory and other techniques
\cite{Ferrari2010, KK2010, Spohn2005, Sasamoto2007}. 
Until a few years ago the analysis had been restricted to the 
totally asymmetric case, $p=0$ or $q=0$, i.e., particles hop only 
in one direction.  

In \cite{TW2008a}, Tracy and Widom succeeded in computing the 
distribution of the particle position for the ASEP with general parameter 
values using the transition probability derived from the Bethe ansatz.  
For recent developments 
see \cite{TW2008b,TW2009a,TW2009b,TW2009c,TW2009d,TW2010a,TW2010b,
          TW2010c,Lee2010}.

Recently their formula was utilized to study the height fluctuations 
of the KPZ equation 
\cite{SS2010a,SS2010b,SS2010c,ACQ2010,CQ2010}.  
By the Cole-Hopf transformation, the KPZ equation is mapped 
to a problem of directed polymer. As noted long time ago,
the $n$ replica partition function of this directed polymer is 
mapped to the problem of attractive $\delta$-Bose gas with $n$ 
particles\cite{Kardar1987,Kardar2007}. This connection has been 
utilized recently in \cite{CLR2010,Dotsenko2010}. By considering 
a generating function of the $n$ replicas, one could reproduce 
the reslults for height fluctuations of the KPZ eqaution. 
See also a related work \cite{OConnell2009}. 

In this note we point out that similar consideration is possible 
for the ASEP. We show that the $n$th exponential moments of current 
in the ASEP can be written as a summation of transition probabilities 
for $k$ particles with $0\leq k \leq n$ of the ASEP. From this one 
could find an expression for current fluctuations of the ASEP. 
Our argument is based on the duality relation of the ASEP found 
by Sch\"utz \cite{Schuetz1997a}.

\section{Duality}\label{sec:duality}
First we consider the ASEP on a finite lattice 
$[L]=\{1,2,\ldots, L\}$
with reflective boundaries. We employ the formulation using the 
quantum spin chain language, see e.g. \cite{Schuetz2000}. 
We set 
\begin{equation}
 s^+ = \begin{bmatrix}
	0 & 1 \\ 0 & 0 
       \end{bmatrix}, \quad
 s^- = \begin{bmatrix}
	0 & 0 \\ 1 & 0
       \end{bmatrix}, \quad
 s^z = \frac12
       \begin{bmatrix}
	1 & 0 \\ 0 & -1
       \end{bmatrix}, \quad
 n = \frac12 -s^z = 
       \begin{bmatrix}
        0 & 0 \\ 0 & 1    
       \end{bmatrix}.
\label{pauli}
\end{equation}
Let us introduce a vector $|0\rangle$ which corresponds to 
the empty system. One can construct an $n$ particle state 
with particle positions at $x_1,\ldots,x_n$ by 
\begin{equation}
 |x_1,\ldots,x_n\rangle = s_{x_1}^- \cdots s_{x_n}^- |0\rangle .
\end{equation}
Here $s_x^-$ means it acts nontrivially only on the space of 
site $x$ as a $2\times 2$ matrix $s^-$ in (\ref{pauli}).
The state of the system can also be specified by a set of 
particle numbers $\eta_x$ on each site $x\in [L]$. Here 
$\eta_x=1$ (resp. $\eta_x=0$) means the site $x$ is occupied
(resp. empty). We sometimes abbreviate 
$\eta=\{\eta_1,\ldots ,\eta_L\}$ and denote the corresponding state 
by $|\eta\rangle$.

Let $P(\eta,t)$ be the probability that the configuration of the system
is $\eta$ at time $t$ and set 
\begin{equation}
 | P\rangle = \sum_{\eta} P(\eta,t) |\eta\rangle 
\end{equation}
where $\sum_{\eta}$ means the summation over all particle configuration.
The time evolution of this is given by 
the master equation, 
\begin{equation}
 \frac{d}{dt} |P\rangle = -H |P\rangle, 
\end{equation}
where the transition rate matrix is given by 
\begin{equation}
 H = -\sum_{j=1}^{L-1} \left[  p s_j^+ s_{j+1}^- + q s_j^- s_{j+1}^+ 
                    -pn_j(1-n_{j+1})-q(1-n_j)n_{j+1} \right]. 
\end{equation}
The transition probability, i.e., the probability that $n$ particles
starting from $y_1,\ldots, y_n$ at time 0 are on sites $x_1,\ldots,x_n$
at time $t$ is written as 
\begin{equation}
 G(x_1,\ldots,x_n;t | y_1,\ldots,y_n;0)
 =
 \langle x_1,\ldots ,x_n | e^{-t H} | y_1,\ldots, y_n \rangle 
\end{equation}
with $\langle x_1,\ldots ,x_n|=\langle 0|s_{x_1}^+ \cdots s_{x_n}^+$. 
We sometimes abbreviate this as $G(\{x\}_n;t|\{y\}_n;0)$ with the understanding
$\{x\}_n=(x_1,\ldots,x_n)$. 

We recall the duality of the ASEP based on a quantum group 
symmetry of the process\cite{Schuetz1997a}. 
Our notation in this article is slightly different from 
\cite{Schuetz1997a}. 
Let us set 
\begin{align}
 X^+ &= \tau^{-L/4+\frac12} \sum_{k=1}^L 
        \tau^{\sum_{j=1}^{k-1} (1-n_j) } s_k^-, \label{Xp}\\
 X^- &= \tau^{-L/4} \sum_{k=1}^L s_k^+ \tau^{\sum_{j=k+1}^L n_j}, 
        \label{Xm}\\
 K   &= \tau^{-\sum_{k=1}^L s_k^z}. \label{K}
\end{align}
They satisfy the $U_q(sl_2)$ algebra \cite{Drinfeld1985,Jimbo1985,PS1990},
\begin{gather}
 K X^+ K^{-1} = \tau X^+, \quad K X^- K^{-1} = \tau^{-1} X^-, \\
 [X^+,X^-] = \frac{K-K^{-1}}{\tau^{1/2}-\tau^{-1/2}}. 
\end{gather}
They commute with $H$, i.e., 
\begin{equation}
\label{HXK}
 [H,X^{\pm}]=[H,K] = 0.
\end{equation}
The last relation reflects the fact that the number of particles 
is conserved in the dynamics. Since the state space is finite 
there is the unique stationary state for each particle number 
$N, 0\leq N \leq L$. Let us denote it by $|N\rangle$. 
It satisfies $H|N\rangle =0$.
When $N=0$ this is nothing but the state $|0\rangle$ with no particle. 
For $N\geq 1$, one can construct $|N\rangle$  by applying $X^+$ 
for $N$ times to $|0\rangle$ as 
\begin{equation}
 |N\rangle 
 = 
 (X^+)^N |0\rangle.            
\label{Nket}
\end{equation}
It is easy to check $H|N\rangle =0$ using (\ref{HXK}) and 
$H|0\rangle =0$.
If we introduce 
\begin{equation}
 \langle N|
 =
 C_N \langle 0|(X^-)^N 
 =
 \sum_{\eta: \sum_{k=1}^L \eta_k = N} \langle \eta|             
\end{equation}
where
\begin{equation}
 C_N = \tau^{\frac{LN}{4}} \frac{(1-\tau)^N}{(1-\tau)\cdots
  (1-\tau^N)},  
\end{equation}
the normalization of a state $|P\rangle $ is written as 
$\langle N|P\rangle=1$.
Using (\ref{Xp}), one sees that the 
normalized version of (\ref{Nket}) is given by  
\begin{equation}
 |N\rangle_{\text{norm}}
 =
 \tau^{-\frac{N(N+1)}{2}}
 \frac{(1-\tau)\cdots (1-\tau^N)}{(1-\tau^{L-N+1})\cdots (1-\tau^L)}
 \sum_{\eta: \sum_{k=1}^L \eta_k = N} 
 \tau^{\sum_{k=1}^L k n_k} |\eta \rangle .
\end{equation}
The average of a quantity $A$ which depends on $\eta_i$'s 
at time $t$ for an initial state
$|I_N\rangle$ with $N$ particle is 
\begin{equation}
 \langle A \rangle_t 
 =
 \langle N| A e^{-Ht} |I_N\rangle. 
\end{equation}
Now let us define 
\begin{equation}
 N_x = \sum_{j=1}^x n_j
\end{equation}
and set
\begin{align}
\label{Qx}
 Q_x &= \tau^{N_x}, \\
 \tilde{Q}_x &= \frac{Q_x-Q_{x-1}}{\tau-1} = \tau^{N_{x-1}}n_x.
\label{Qt}
\end{align}
One can verify
\begin{lem}\label{lem:XNQ}
\begin{equation}
 [(X^-)^N, Q_x]
 =
 (\tau^N-1)Q_x X_x^- (X^-)^{N-1} 
\label{XNQ}
\end{equation}
where 
\begin{equation}
 X_x^- = \tau^{-L/4}\sum_{k=1}^x s_k^+ \tau^{\sum_{j=k+1}^L n_j}.
\end{equation} 
\end{lem}
\begin{proofOF}{Lemma \ref{lem:XNQ}}
The $N=1$ case is easily checked by using the following relations 
\begin{equation}
 \tau^n s^+ = s^+, ~~ s^- \tau^n = \tau s^+, ~~
 \tau^n s^- = \tau s^-, ~~ s^- \tau^n = s^-. 
\label{tau_s}
\end{equation}
Next assume (\ref{XNQ}) is true for $N$. We want to see 
(\ref{XNQ}) holds for $N+1$. We start from 
\begin{equation}
 [(X^-)^{N+1},Q_x]
 =
 [(X^-)^N,Q_x]X^- + (X^-)^N [X^-,Q_x].
\end{equation}
One uses (\ref{XNQ}) for $N=1$ and $N$ to get 
\begin{align}
 &[(X^-)^{N+1},Q_x] \notag\\
 &=
 (\tau^N -1)Q_x X_x^- (X^-)^N 
 +
 (\tau^N -1)(\tau-1)Q_x X_x^- (X^-)^{N-1} X_x^-
 +
 (\tau-1)Q_x (X^-)^N X_x^- .
\end{align}
Comparing this with RHS of (\ref{XNQ}) for $N+1$, it is 
enough to show
\begin{equation}
 (X^--X_x^-)(X^-)^{N-1}X_x^-
 =
 \tau^N X_x^- (X^-)^N (X^- -X_x^-). 
\end{equation}
To verify this, one can check the $N=1$ case by using 
(\ref{Xm}),(\ref{tau_s}) and then use mathematical induction.

\end{proofOF}
Applying this lemma, we see, when $x_i$'s are distinct, 
\begin{equation}
 \langle N| \tilde{Q}_{x_1} \cdots \tilde{Q}_{x_n}
 =
 C_{N,n} \langle x_1,\ldots ,x_n|(X^-)^{N-n} 
\label{NQ}
\end{equation}
where
\begin{equation}
 C_{N,n} =  \frac{\tau^{\frac14 (N-n)L }  (1-\tau)^{N-n} }
      {(1-\tau)\cdots (1-\tau^{N-n})}.
\end{equation}
Notice $C_N = C_{N,0}$. Using this one can show 
\begin{prop} [\cite{Schuetz1997a}]
 When $x_i$'s are distinct, it holds 
\begin{equation}
 \langle \tilde{Q}_{x_1} \cdots \tilde{Q}_{x_n}\rangle_t
 =
 \sum_{1\leq y_1 < \ldots < y_n \leq L} 
 G(x_1,\ldots,x_n;t | y_1,\ldots, y_n;0) 
 \langle \tilde{Q}_{y_1} \cdots \tilde{Q}_{y_n}\rangle_0.
\end{equation}
\label{duality}
\end{prop}

\begin{proofOF}{Proposition \ref{duality}}
This is seen as follows:
\begin{align}
 \text{LHS}
 &=
 \langle N| \tilde{Q}_{x_1} \cdots \tilde{Q}_{x_n} e^{-H t} |I_N\rangle
 \notag\\
 &=
 C_{N,n} \langle x_1,\ldots ,x_n|(X^-)^{N-n} e^{-H t}|I_N\rangle \notag\\
 &=
 C_{N,n} \langle x_1,\ldots ,x_n| e^{-H t} (X^-)^{N-n} |I_N\rangle \notag\\
 &= 
 \sum_{1\leq y_1 < \ldots < y_n \leq L} 
 \langle x_1,\ldots ,x_n| e^{-H t} |y_1,\ldots, y_n\rangle
 \cdot C_{N,n} \langle y_1,\ldots ,y_n| (X^-)^{N-n} |I_N\rangle \notag\\
 &=
 \sum_{1\leq y_1 < \ldots < y_n \leq L} 
 \langle x_1,\ldots ,x_n| e^{-H t} |y_1,\ldots, y_n\rangle
 \langle N| \tilde{Q}_{y_1} \cdots \tilde{Q}_{y_n} |I_N\rangle 
 =
 \text{RHS}. 
\end{align}
Here $C_{N,n}$ is the constant appearing in (\ref{NQ}). In the third 
equality we used (\ref{HXK}) and in the forth equality we used 
the fact that 
$\sum_{ 1\leq y_1 < \ldots < y_n \leq L} 
| y_1,\ldots ,y_n \rangle \langle y_1,\ldots ,y_n|$ 
acts as an identity in the subspace with $n$ particles. 
\end{proofOF}
This is a generalization of the well known duality for the 
symmetric simple exclusion process
\cite{Liggett1985}. The computation of $k$ point correlation 
functions of (\ref{Qt}) is reduced to the $k$ particle problem. 

To study the exponential moments of currents, we need a formula
when $x_i$'s are equal. It turns out that 
the quantity can not be written as a summation of only $n$-particle 
transition probability but as a sum of $k$ particle ones 
for all $k (\leq n)$. The result is
\begin{prop}\label{Prop:Qn}
\begin{equation}
 \label{Qn}
 \langle Q_x^n \rangle_t
 =
 \sum_{k=0}^n (\tau^n-1)\cdots(\tau^{n-k+1}-1) 
 \sum_{1\leq x_1 < \ldots < x_k \leq x} 
 \sum_{1\leq y_1 < \ldots < y_k \leq L} 
 G(\{x\}_k;t |\{y\}_k ;0) 
 \langle \tilde{Q}_{y_1} \cdots \tilde{Q}_{y_k}\rangle_0.
\end{equation}
\end{prop}

To derive this we need a few lemmas. One first shows
\begin{lem}\label{duality_n}
\begin{equation}
 \langle N | \tilde{Q}_x^2
 =
 (\tau-1) \sum_{j=1}^{x-1} \langle N| \tilde{Q}_j \tilde{Q}_x
 + \langle N | \tilde{Q}_x 
\label{Q2}
\end{equation} 
\end{lem}
\begin{proofOF}{Lemma \ref{duality_n}}
First one computes  
\begin{equation}
 \text{LHS}
 =
 C_N \langle 0|(X^-)^N \tilde{Q}_x^2 
 =
 C_N \langle 0| [(X^-)^N, \tilde{Q}_x^2]
 =
 C_N \langle 0| [(X^-)^N, \tilde{Q}_x] \tilde{Q}_x .
\end{equation}
Using Lemma \ref{lem:XNQ}, this becomes
\begin{equation}
 C_N \frac{\tau^N-1}{\tau-1}
 \langle 0|(Q_x X_x^- -Q_{x-1} X_{x-1}^-)
 (X^-)^{N-1} \tilde{Q}_x.
\end{equation} 
Rewriting 
$(X^-)^{N-1} \tilde{Q}_x = 
 [(X^-)^{N-1}, \tilde{Q}_x ] + \tilde{Q}_x (X^-)^{N-1}$,
applying Lemma \ref{lem:XNQ} again and using (\ref{Xm}), it is 
\begin{align}
 &\tau^{-L/2}\frac{\tau^N-1}{\tau-1}
 \left\{ \frac{\tau^{N-1}-1}{\tau-1}\langle 0|
         \left( (\tau-1)\sum_{j=1}^{x-1} s_j^+ s_x^+  \right)
	 (X^-)^{N-2} + \langle 0|s_x^+ \tilde{Q}_x (X^-)^{N-1}     
   \right\} \notag\\
 &= 
 (\tau-1)C_{N,2} \sum_{j=1}^{x-1}\langle j,x|(X^-)^{N-2} 
 +
 C_{N,1}\langle x|(X^-)^{N-1}.  
\end{align}  
By (\ref{NQ}) this is the RHS of (\ref{Q2}).
\end{proofOF}
Let us define for a fixed $x$
\begin{equation}
 B_n = \langle N| \sum_{1\leq x_1 < \cdots <x_n \leq x}
       \tilde{Q}_{x_1} \cdots \tilde{Q}_{x_n} 
\end{equation}
Then we have
\begin{lem}\label{Lem:BQ}
\begin{equation}
 B_n \sum_{j=1}^x \tilde{Q}_j 
 =
 \frac{1-\tau^{n+1}}{1-\tau}B_{n+1}
 +
 \frac{1-\tau^n}{1-\tau}B_n
\label{BQ}
\end{equation}
 \end{lem}
\begin{proofOF}{Lemma \ref{Lem:BQ}}
First we see
\begin{equation}
 B_n \sum_{j=1}^x \tilde{Q}_j 
 =
 (n+1) B_{n+1} + \sum_{i=1}^n B_{n,i}
\label{BAim}
\end{equation}
where
\begin{equation}
 B_{n,i} 
 = 
 \sum_{1\leq x_1 < \cdots < x_l \leq x}
 \langle N| \tilde{Q}_{x_1} \cdots \tilde{Q}_{x_i}^2 \cdots 
 \tilde{Q}_{x_n},
\end{equation}
This can be rewritten in terms of $B_n,B_{n+1}$ only as 
\begin{equation}
 B_{n,i}
 =
 (\tau^i-1)B_{n+1}+\tau^{i-1}B_n. 
\label{Aim}
\end{equation} 
This is seen as follows. 
Using (\ref{Q2}), one has 
\begin{align}
 B_{n,1} &= (\tau-1)B_{n+1} + B_n, \\ 
 B_{n,i} &= (\tau-1)(B_{n,1}+\cdots + B_{n,i-1})
            + i(\tau-1)B_{n+1} + B_n. 
\end{align}
Suppose (\ref{Aim}) is correct for $1,2,\ldots ,i-1$. Then
one gets (\ref{Aim}) for $i$ by mathematical induction. 
Plugging (\ref{Aim}) into (\ref{BAim}), we get (\ref{BQ}). 
\end{proofOF}

Now using lemmas \ref{duality_n},\ref{Lem:BQ}, it is not difficult 
to show proposition \ref{Prop:Qn} by mathematical induction.

\section{Step Markov initial condition for the ASEP on $\mathbb{Z}$}
To consider the ASEP on $\mathbb{Z}$, we first put the ASEP with 
reflective boundaries of size $2L+1$ on $\{-L,-L+1,\ldots,L-1,L\}$. 
Then by taking the $L\to\infty$ limit in (\ref{Qn}), we have, 
for the ASEP on $\mathbb{Z}$,  
\begin{align}
 \langle Q_x^n \rangle_t
 &=
 \sum_{k=0}^n (\tau^n-1)\cdots(\tau^{n-k+1}-1) \notag\\
 &\quad \times 
\sum_{-\infty < x_1 < \ldots < x_k \leq x} 
 \sum_{-\infty < y_1 < \ldots < y_k < \infty} 
 G(\{x\}_k;t | \{y\}_k;0) 
 \langle \tilde{Q}_{y_1} \cdots \tilde{Q}_{y_k}\rangle_0.
\label{QnG}
\end{align}
Here $G(\{x\}_k;t|\{y\}_k;0)$ is the transition probability of 
the ASEP on $\mathbb{Z}$ and $\tilde{Q}_x$ is defined by 
(\ref{Qx}),(\ref{Qt}) with 
\begin{equation}
 N_x = \sum_{j=-\infty}^x n_j. 
\end{equation}
Let us assume $q>p$ from now on. 
For the summation in (\ref{QnG}) to converge, we assume in the sequel
of the paper that there are no particles far to the left and there are 
enough many particles far to the right.
With this in mind we state the formula as 
\begin{prop}\label{prop:QnGc}
The $n$th moment of $Q_x$ is written as 
\begin{equation}
 \langle Q_x^n \rangle_t
 =
 \sum_{k=0}^n (\tau^n-1)\cdots(\tau^{n-k+1}-1) c_k
\label{QnG2}
\end{equation}
with 
\begin{equation}
 c_k = \sum_{-\infty < x_1 < \ldots < x_k \leq x} 
 \sum_{-\infty < y_1 < \ldots < y_k < \infty} 
 G(\{x\}_k;t | \{y\}_k;0)
 \langle \tilde{Q}_{y_1} \cdots \tilde{Q}_{y_k}\rangle_0.
\end{equation}
\end{prop}
Let $N_t(x)$ be the integrated current at bond between sites 
$x$ and $x+1$, i.e., the number of particles which hop from 
$x+1$ to $x$ minus the number of particles which hop from 
$x$ to $x+1$ up to time $t$. Notice that when we consider 
the initial condition such that $\eta_x=0,x\leq 0$, then 
$N_t(x)$ is the number of particles on sites $\leq x$ and 
hence $\langle Q_x^n\rangle_t = \langle \tau^{n N_t(x)} \rangle$ 
for $x\leq 0$.
Hence the quantity in Proposition \ref{prop:QnGc} is the 
same as the exponential moment of the current of the ASEP. 
For $x>0$ and more general initial conditions, one has to modify 
the relation between $Q_x$ and $N_t(x)$ to incorporate the 
initial configuration of particles. 
  
For the ASEP on $\mathbb{Z}$, the transition probability 
is written as \cite{Schuetz1997b,TW2008a}
\begin{equation}
 G(\{x\}_k;t|\{y\}_k;0)
 =
 \sum_{\sigma\in S_k}
 \int_{C_R} \cdots \int_{C_R} d\xi_1\cdots d\xi_k
 A_{\sigma} \prod_i \xi_{\sigma(i)}^{x_i-y_{\sigma(i)}-1}
 e^{\sum_i \epsilon(\xi_i) t}
\label{Green}
\end{equation}
where $S_k$ is a set of all permutations of order $k$, 
$\epsilon(\xi) = p/\xi + q \xi -1$ and
\begin{equation}
 A_{\sigma} 
 = 
 \text{sgn} \sigma 
 \frac{\prod_{i<j} (p+q \xi_{\sigma(i)}\xi_{\sigma(j)} -\xi_{\sigma(i)})}
      {\prod_{i<j} (p+q \xi_i \xi_j -\xi_i)}.
\end{equation}
$C_R$ is a contour enclosing the origin anticlockwise with 
a radius large enough that all the poles in $A_\sigma$ are included 
in $C_R$. In \cite{TW2008a}, the contour was taken to be a small one, 
but one can simply use the transformation $\xi \to 1/\xi$ to switch 
to a large contour.  

Suppose we substitute this representation into (\ref{QnG2}). 
Taking $R>1$, one has $|\xi_i|>1, 1\leq i\leq k$ so that the 
summation over $x$ can be performed as 
\begin{equation}
 \sum_{-\infty<x_1<\ldots <x_k\leq x}
 \xi_{\sigma(1)}^{x_1} \cdots \xi_{\sigma(k)}^{x_k} 
 =
 \frac{(\xi_1 \cdots \xi_k)^{x+1}}
 {(\xi_{\sigma(1)}-1)\cdots (\xi_{\sigma(1)}\cdots \xi_{\sigma(k)}-1)}.
\label{sum_x}
\end{equation}
By using (\ref{Green}),(\ref{sum_x}) and a formula given in 
\cite{TW2008a}, 
\begin{equation}
 \sum_{\sigma\in S_k} \text{sgn}\sigma
 \frac{\prod_{i<j} (p+q \xi_{\sigma(i)}\xi_{\sigma(j)} -\xi_{\sigma(i)})}
      {(\xi_{\sigma(1)}-1)\cdots (\xi_{\sigma(1)}\cdots \xi_{\sigma(k)}-1)} 
 =
 (-1)^k q^{\frac12 k(k-1)} \frac{\prod_{i<j}(\xi_j-\xi_i)}{\prod_i (1-\xi_i)},
\end{equation}
one arrives at 
\begin{thm}\label{thm:cQ}
The $n$th moment of $Q_x$ is written as (\ref{QnG2})
with 
\begin{align}
 c_k 
 &=
 (-1)^k q^{\frac12 k(k-1)}
 \int_{C_R}\cdots \int_{C_R} d\xi_1 \ldots d\xi_k
 \prod_{i<j}\frac{\xi_j-\xi_i}{p+q\xi_i \xi_j-\xi_i}
 \prod_i \frac{ \xi_i^xe^{\epsilon(\xi_i)t}}{1-\xi_i}\notag\\
&\quad \times
  \sum_{-\infty < y_1 < \ldots < y_k < \infty} 
  \frac{ \langle \tilde{Q}_{y_1} \cdots \tilde{Q}_{y_k}\rangle_0}
      {\xi_1^{y_1} \cdots \xi_k^{y_k}}.
\label{GQ}
\end{align}
\end{thm}

\noindent
In this expression, the dependence on the initial condition is 
clearly separated. It is straightforward to check whether the 
summation over $y$ can be taken for a given initial condition. 

From (\ref{GQ}) one guesses that the distribution function of the
current is given by (again for a special case where $\eta_x=0,x\leq 0$
initially) 
\begin{equation}
 \mathbb{P}\left[ N_t(x) \geq m\right]
 =
 (-1)^m \tau^{\frac12 m(m-1)} 
 \sum_{k\geq 0}\tau^{(1-m)k} \binom{k-1}{k-m}_\tau c_k, ~x\leq 0 
\label{PNt}
\end{equation}
where $\binom{N}{n}_\tau$ is $\tau$-binomial coefficient difined as
$\binom{N}{n}_\tau 
= 
\frac{(1-\tau^N)\cdots (1-\tau^{N-n+1})}{(1-\tau)\cdots (1-\tau^n)}$.
In fact following the argument in \cite{TW2010a}, one sees that
(\ref{PNt}) leads to the expression, 
\begin{equation}
 \langle e^{\lambda N_t(x)} \rangle
 =
 \sum_{k=0}^{\infty} \tau^{-\frac12 k(k-1)} e^{\lambda k}
 \prod_{j=0}^{k-1} (1-e^{-\lambda}\tau^j) \cdot c_k.
\label{elN}
\end{equation}
When $\lambda=n \log \tau, n\in\{1,2,\ldots\}$, the summation over $k$ 
is terminated at $k=n$ and this reduces to (\ref{QnG2}). 
Of course to go in the opposite way from (\ref{QnG2}) to (\ref{elN}), 
there is a question of analytic continuation, an infamous problem in 
replica theory. We do not
discuss it here but just mention that it looks natural to expect 
(\ref{elN}) from (\ref{QnG2}). 

To further proceed, we need to take the summation in (\ref{GQ}).
As explained in \cite{TW2010c}, there are not many examples for which 
this has been done. Here we will give a generalization by observing 
\begin{lem} \label{lem:sumQ}
 Suppose $\langle \tilde{Q}_{y_1} \cdots \tilde{Q}_{y_k}\rangle_0$
 has the form, 
\begin{equation}
 \langle \tilde{Q}_{y_1} \cdots \tilde{Q}_{y_k}\rangle_0
 =
 a_k 1_{y_1\geq 1} \prod_{i=1}^k g_i(y_i-y_{i-1}-1)
\label{Qg}
\end{equation}
 with the convection $y_0=0$. Here $a_k$ does not depend on $y_i$
 and the function $g_i,1\leq i\leq k$ is assumed to be 
 such that $\sum_{y=0}^{\infty} \frac{g_i(y)}{\xi^{y+1}}$ converges 
 for $|\xi|$ large enough. Then
 \begin{equation}
  \sum_{1 \leq y_1 < \ldots < y_k < \infty} 
 \frac{ \langle \tilde{Q}_{y_1} \cdots \tilde{Q}_{y_k}\rangle_0}
      {\xi_1^{y_1} \cdots \xi_k^{y_k}}
 = a_k
 \prod_{i=1}^k \sum_{y_i=0}^{\infty}
 \frac{g_i(y_i)}{(\xi_i \cdots \xi_k)^{y_i+1}}. 
\end{equation}
\end{lem}
\begin{proofOF}{Lemma \ref{lem:sumQ}}
 Due to $1_{y_1\geq 1}$ in (\ref{Qg}), the summation in (\ref{GQ}) 
 can be replaced by 
 $\sum_{1\leq y_1 < y_2  < \ldots < y_k}$. By shifting $y_i\to y_i-1$,
 the LHS reads
 \begin{equation}
  a_k
  \sum_{y_1=0}^{\infty} \sum_{y_2=y_1+1}^{\infty} \cdots
  \sum_{y_k=y_{k-1}+1}^{\infty}
  \frac{g_1(y_1)}{\xi_1^{y_1+1}}
  \frac{g_2(y_2-y_1-1)}{\xi_2^{y_2+1}}
  \frac{g_k(y_k-y_{k-1}-1)}{\xi_k^{y_k+1}}. 
 \end{equation}
 One further makes a change of variable, 
 $y_i\to y_1 + (y_2+1) +\cdots (y_i+1)$, 
 to get the RHS. 
\end{proofOF}

Let us consider the initial condition in which there is a particle 
at the origin and the particle occupation on 
$\mathbb{Z}_+=\{1,2,\ldots\}$ is described as a Markov process with 
the 2$\times$2 transition matrix,
\begin{equation}
 A = \begin{bmatrix}
      1-\rho & 1-\mu \\ \rho & \mu
     \end{bmatrix}
\label{A}
\end{equation}
where $0<\rho,\mu<1$. This means that if the site $x$($\geq 1$) is 
empty then the site $x+1$ is empty (resp. occupied) with probability 
$1-\rho$ (resp. $\rho$) and if the site $x$($\geq 1$) is 
occupied then the site $x+1$ is empty (resp. occupied) with probability 
$1-\mu$ (resp. $\mu$). We call this the step Markov initial condition. 
When $\rho=\mu$, each site $x\geq 1$ is
independent and this becomes the step Bernoulli initial conditions 
\cite{TW2009b}
except that there is a particle at the origin. 
When the site $x$ is occupied, let us denote by $p_L(m)$ the 
the probability that there is a particle at site $x+L+1$ and 
that there are $m$ particles on sites from $x+1$ up to $x+L$. We set
\begin{equation}
 w_L(\zeta) = \sum_{m=0}^L p_L(m) \zeta^m.  
\label{w}
\end{equation}
Notice, if one defines
\begin{equation}
 Z_L(\zeta) = \langle 1| \left( A \begin{bmatrix}
			      1 & 0 \\ 0 & \zeta
			     \end{bmatrix} \right)^L A |1\rangle ,~~
 |1\rangle = ~^t(0,1), 
\label{Z}
\end{equation}
this is written as $w_L(\zeta) = Z_L(\zeta)/Z_L(1)$.

Let $y_0=0, 1\leq y_1<y_2<\cdots < y_k$. 
Suppose there are particles on $y_i, 1\leq i\leq k$ and that  
there are $m_i$ particles on sites between $y_{i-1}+1$ and 
$y_i-1,~~1\leq i \leq k$. 
This happens with probability 
\begin{equation}
  \prod_{i=1}^k p_{y_i-y_{i-1}-1}(m_i)
\label{prod_p}
\end{equation}
due to the Markov properties of the measure. We also have
\begin{equation}
 \tilde{Q}_{y_1}\cdots \tilde{Q}_{y_k}
 =
 \prod_{i=1}^k \tau^{m_1+\ldots +m_i +i}
 =
 \tau^{\frac12 k(k+1)+\sum_{i=1}^k (k-i+1)m_i}.
\end{equation}
Hence 
\begin{align}
 \langle \tilde{Q}_{y_1} \cdots \tilde{Q}_{y_k}\rangle_0
 &=
 \tau^{\frac12 k(k+1)} 
 \prod_{i=1}^k \sum_{m_i=0}^{y_i-y_{i-1}-1} p_{y_i-y_{i-1}-1}(m_i)
 \tau^{(k-i+1)m_i} \notag\\ 
 &=
 \tau^{\frac12 k(k+1)} 
 \prod_{i=1}^k w_{y_i-y_{i-1}-1}(\tau^{k-i+1}).
\end{align}
This is exactly the form in Lemma \ref{lem:sumQ} with 
$g_i(y) = w_y(\tau^{k-i+1})$ and hence 
\begin{align}
&\quad
 \sum_{-\infty < y_1 < \ldots < y_k < \infty} 
 \frac{ \langle \tilde{Q}_{y_1} \cdots \tilde{Q}_{y_k}\rangle_0}
      {\xi_1^{y_1} \cdots \xi_k^{y_k}}
=
 \tau^{\frac12 k(k+1)} 
 \prod_{i=1}^k \sum_{y_i=0}^{\infty}
 \frac{w_{y_i}(\tau^{k-i+1})}{(\xi_i \cdots \xi_k)^{y_i+1}}. 
\label{sumQw}
\end{align}

Combining Theorem \ref{thm:cQ} and Lemma \ref{lem:sumQ}, we summarize our
result as 
\begin{thm} \label{thm:stepM}
For the step Markov initial condition described around (\ref{A}),
the $n$th moment of $Q_x$ is given by (\ref{QnG2})
with
\begin{align}
 c_k 
 &=
 (-1)^k q^{\frac12 k(k-1)} \tau^{\frac12 k(k-1)}
 \int_{C_R}\cdots \int_{C_R} d\xi_1 \ldots d\xi_k
 \prod_{i<j}\frac{\xi_j-\xi_i}{p+q\xi_i \xi_j-\xi_i}
 \prod_i \frac{ \xi_i^xe^{\epsilon(\xi_i)t}}{1-\xi_i}\notag\\
&\quad \times
 \prod_{i=1}^k \sum_{y_i=0}^{\infty}
 \frac{w_{y_i}(\tau^{k-i+1})}{(\xi_i \cdots \xi_k)^{y_i+1}}.
\label{c_w}
\end{align}
\end{thm}

\noindent
It may still be in general nontrivial to take the summation over 
$y_i$ in (\ref{c_w}) explicitly but the point here is that 
the summations over $y_i$'s are now separated and hence one may 
be able to do asymptotics using this expression with 
(\ref{w}),(\ref{Z}). 

In the parallel way, 
one can also treat the initial condition in which there is no particle 
at the origin and the particle occupation on 
$\mathbb{Z}_+=\{1,2,\ldots\}$ is described as a Markov process with 
the transition matrix (\ref{A}). The only difference is that 
the first product $p_{y_1-1}(m_1)$ is replaced by 
$p^{(0)}_{y_1-1}(m_1)$ in (\ref{prod_p}) where $p^{(0)}_L(m)$ is the 
the probability that, when the site $x$ is empty, 
there is a particle at site $x+L+1$ and that there are $m$ particles 
on sites from $x+1$ up to $x+L$. We also define 
$w_L^{(0)}(\zeta)$ accordingly. Then the net change in (\ref{c_w}) 
is that one replaces $\tau^{\frac12 k(k+1)}$ by $\tau^{\frac12 k(k-1)}$
and $w_{y_1}(\tau^k)$ by $w^{(0)}_{y_1}(\tau^k)$ respectively. 

When $\mu=\rho$, the measure (assuming there is no particle at the 
origin) becomes the step Bernoulli initial conditions. 
For this case, one has 
\begin{align}
 p_L(m) &= \rho \binom{L}{m} \rho^m (1-\rho)^{L -m}, \\
 w_L(\zeta) &= \sum_{m=0}^L p_L(m) \zeta^m =
 \rho (1-\rho(1-\zeta))^L.  
\end{align}
Applying Lemma \ref{lem:sumQ} with
$a_k=\rho^k,g_i(y)=(1-\rho(1-\zeta))^y$,
one finds 
\begin{equation}
 \sum_{y_i=0}^{\infty}\frac{g_i(y_i)}{(\xi_i \cdots \xi_k)^{y_i+1}} 
 =
 \frac{1}{\xi_i\cdots \xi_k-1+\rho(1-\tau^{k-i+1})}.
\end{equation} 
The final result is given by 
\begin{equation}
\sum_{-\infty < y_1 < \ldots < y_k < \infty} 
 \frac{ \langle \tilde{Q}_{y_1} \cdots \tilde{Q}_{y_k}\rangle_0}
      {\xi_1^{y_1} \cdots \xi_k^{y_k}}
=
\tau^{\frac12 k(k-1)} \rho^k 
\prod_{i=1}^k \frac{1}{\xi_i\cdots \xi_k-1+\rho(1-\tau^{k-i+1})}.
\end{equation}
This agrees with the expression given in \cite{TW2009d}. 

Next we consider the $m$-periodic initial conditions in which  
particles start from the sites $1,m+1,2m+1,\ldots$ \cite{Lee2010}. 
In this case one has
\begin{equation}
 \tilde{Q}_y = 1_{\stackrel{y\geq 1}{y\equiv 1}}
 \tau^{\frac{y-1}{m}}
\end{equation}
where $\equiv$ means the modulo $m$. 
Hence 
\begin{align}
 \sum_{-\infty < y_1 < \ldots < y_k < \infty} 
 \frac{ \langle \tilde{Q}_{y_1} \cdots \tilde{Q}_{y_k}\rangle_0}
      {\xi_1^{y_1} \cdots \xi_k^{y_k}}
=
\sum_{\stackrel{y_1=1}{y_1\equiv 1}}^{\infty}
\frac{\tau^{\frac{y_1-1}{m}}}{\xi_1^{y_1}}
\sum_{\stackrel{y_2=y_1+m}{y_2\equiv 1}}^{\infty}
\frac{\tau^{\frac{y_2-1}{m}}}{\xi_2^{y_2}}
\cdots
\sum_{\stackrel{y_k=y_{k-1}+m}{y_k\equiv 1}}^{\infty}
\frac{\tau^{\frac{y_k-1}{m}}}{\xi_k^{y_k}}. 
\end{align}
This is not exactly the form treated by Lemma \ref{lem:sumQ}. 
But if we make a change of variable,
$y_i = m \tilde{y}_i -m+1,~ 1\leq i\leq k$, then 
\begin{align}
 \frac{ \langle \tilde{Q}_{y_1} \cdots \tilde{Q}_{y_k}\rangle_0}
      {\xi_1^{y_1} \cdots \xi_k^{y_k}}
=
\tau^{\frac12 k(k-1)}1_{\tilde{y}_1 \geq 1}
\prod_{i=1}^k \tau^{(k-i+1)(\tilde{y}_i-\tilde{y}_{i-1}-1)}. 
\end{align}
Now one can apply Lemma \ref{lem:sumQ} with 
$g_i(y) = \tau^{(k+i-1)y}$ to obtain 
\begin{equation}
\sum_{-\infty < y_1 < \ldots < y_k < \infty} 
 \frac{ \langle \tilde{Q}_{y_1} \cdots \tilde{Q}_{y_k}\rangle_0}
      {\xi_1^{y_1} \cdots \xi_k^{y_k}}
=
\tau^{\frac12 k(k-1)} (\xi_1 \cdots \xi_k)^{m-1}
\prod_{i=1}^k \frac{1}{\xi_i^m\cdots \xi_k^m-\tau^{k-i+1}}.
\end{equation}
This agrees with the expression given in \cite{TW2010a,Lee2010}. 

\medskip
\noindent
\textbf{Acknowledgements.} We are grateful to H. Spohn for useful 
discussions and comments. T.S. acknowledges the support 
from KAKENHI (22740054)

\providecommand{\bysame}{\leavevmode\hbox to3em{\hrulefill}\thinspace}
\providecommand{\MR}{\relax\ifhmode\unskip\space\fi MR }
\providecommand{\MRhref}[2]{%
  \href{http://www.ams.org/mathscinet-getitem?mr=#1}{#2}
}
\providecommand{\href}[2]{#2}


\end{document}